\documentclass[entropy,article,accept,moreauthors,10pt,a4paper]{mdpi}

\usepackage[english]{babel}
\usepackage{xcolor}
\usepackage{pgf}
\usepackage{geometry}
\usepackage{graphicx,psfrag,etoolbox}
\usepackage{dsfont}

\newtheorem{thm}{Theorem}[section]
\newtheorem{cor}{Corollary}[section]

\newcommand{\abs}[1]{\left\vert#1\right\vert}
\newcommand{\ass}{\stackrel{\textup{\tiny def}}{=}}

\firstpage{1}
\articlenumber{x}
\doinum{10.3390/------}
\pubvolume{xx}
\pubyear{2017}
\copyrightyear{2017}
\externaleditor{Academic Editor: name}
\history{Received: date; Accepted: date; Published: date}

\Title{On the significance of the quantum mechanical covariance
  matrix}

\Author{Avishy Carmi $^{\dagger}$ and Eliahu Cohen $^{\ddagger}$}

\AuthorNames{Avishy Carmi and Eliahu Cohen}

\address{%
$^{\dagger}$ \quad Center for Quantum Information Science and
Technology \& Faculty of Engineering Sciences \\ Ben-Gurion University
of the Negev, Beersheba 8410501, Israel. \\
$^{\ddagger}$ \quad Physics Department, Centre for Research in Photonics,
  University of Ottawa, Advanced Research Complex, 25 Templeton,
  Ottawa ON Canada, K1N 6N5.}

\abstract{The characterization of quantum correlations, being stronger
  than classical, yet weaker than those appearing in non-signaling models, still poses many riddles. In this work we show
that the extent of binary correlations in a general class of
nonlocal theories can be characterized by the existence of a certain
covariance matrix. The set of quantum realizable two-point
correlators in the bipartite case then arises
from a subtle restriction on the structure of this general covariance
matrix. We also identify a class of theories whose covariance does not
have neither a quantum nor an ``almost quantum'' origin, but which
nevertheless produce the accessible two-point quantum mechanical
correlators. Our approach leads to richer Bell-type inequalities in which the extent of
nonlocality is intimately related to a non-additive entropic
measure. In particular, it suggests that the Tsallis entropy with parameter
$q=1/2$ is a natural operational measure of non-classicality.
Moreover, when generalizing this covariance
matrix we find novel characterizations of the quantum mechanical set
of correlators in multipartite scenarios. All these predictions might be
experimentally validated when adding weak measurements to the
conventional Bell test (without adding postselection).}

\begin{document}

\section{Introduction}

The extent of nonlocality is commonly determined by a set of
correlations. In the simplest bipartite scenario, the four two-point correlators
$c_1$, $c_2$, $c_3$ and $c_4$, corresponding to the four
pairs of possible outcomes of Alice and Bob, may render the theory
classical, quantum, or stronger-than-quantum. In the following paper
we plan to tell the richer story provided by a certain covariance
matrix to be presented in the next section. This matrix, which may be defined in any
statistical theory, implies a bound on two-point correlators analogous to that
of quantum mechanics. We thus prove that all potential theories having a covariance structure similar to that of quantum mechanics have a similar set of realizable correlators. Interestingly, this is yet less than the structure imposed by quantum and almost quantum theories \cite{AQ}.
These results cast light on the origin of quantum
correlations, they suggest that other hypothetical theories might
exist which are indistinguishable at the level of correlations from
both quantum and almost quantum theories. In this
  sense, our work can be seen as part of the efforts (see
  e.g. \cite{PR, NPA, Infocause, ML, SuperQ, Bellnonlocality:14, new, ours}) to achieve better qualitative and quantitative understanding of quantum nonlocality.

This paper has two main parts. The first is general and does not rely on the quantum-mechanical
formalism to characterize nonlocality. The second, which builds on these general results,
assumes a quantum structure to derive new bounds on bipartite and
tripartite two-point correlators.

Among the preceding papers in this area, there are mainly two other works
where covariance and second moment matrices, different from the ones
considered here, are used for characterizing quantum mechanical
correlations and probability distributions- The NPA test \cite{NPA},
which significantly extends the approach previously employed in
\cite{L88}. We note the following primary difference between these
works and the paper at hand. While the positive semi-definiteness property
plays a role in both, the particular covariance
here leads to the identification of fundamental relations between the
entries in this matrix. These relations alone are shown to govern the
set of realizable binary bipartite correlators not only in quantum mechanics but in
any nonlocal theory, and to imply new tighter bounds on this set.

\section{Covariance-based certificate of nonlocality}

We shall restrict ourselves for the moment to the Bell-CHSH \cite{Bell,CHSH} setup where two experimenters perform measurements with
one of their measurement devices. Alice measures using either her device
$0$ or device $1$, and similarly Bob measures using either his
device $0$ or device $1$. Both Alice's outcome $a_i$ when she measured
using device $i$ and Bob's outcome $b_j$ when he measured using
device $j$ may either be $1$ or $-1$. We consider the products
$x_{1+i+2j} = a_i b_j$ in different experiments where Alice and Bob
used the pair of devices $i,j$. In a local hidden
  variables theory the Bell--CHSH inequality, $\abs{E[x_1] + E[x_2] +
  E[x_3] - E[x_4]} \leq 2$, holds~\cite{CHSH}.



Suppose now there exists a covariance matrix underlying the products
$x_1,\ldots, x_4$. This $4 \times 4$ matrix is defined as
\begin{equation}
\label{eq:cv1}
\mathcal{C} \ass \mathcal{M} - VV^T,
\end{equation}
where $\mathcal{M}$ is a positive semi-definite second moment matrix
whose diagonal entries all equal $1$, and $V^T = [c_1,\ldots,c_4]$ is
the vector of two-point correlators. If the product $x_i$ is a realization of
the random variable $X_i$ then $\mathcal{M}_{ij} \ass E[X_i X_j]$ and $c_i \ass
E[X_i]$, and if it is associated with an operator $X_i$  (as in
quantum mechanics) then $\mathcal{M}_{ij} \ass \frac{1}{2} \langle \{X_i, X_j\} \rangle$ and
$c_i \ass \langle X_i \rangle$, where $\{X_i, X_j\} \ass X_i X_j + X_j
X_i$ is the anti-commutator. The covariance is by construction real,
symmetric and positive semi-definite.

But even without specifying how the covariance is evaluated,
$\mathcal{C} \succeq 0$ (which means hereinafter that $\mathcal{C}$ is positive semidefinite) may be understood as an algebraic constraint on the
vector of correlators that allows a covariance matrix to be defined in
the underlying theory. In particular,

\begin{equation}
\label{eq:ineq1}
VV^T \preceq \mathcal{M},
\end{equation}
which geometrically means that $V$ is confined to the ellipsoid described
by $\mathcal{M}$. For example, a theory having no constraints
whatsoever on the correlators may have $\mathcal{M} = VV^T$. The PR-box
is one such theory. It is worth noting that in the language of
\cite{CarmiMoskovich} the left-hand side in
\eqref{eq:ineq1} is a Fisher information matrix associated with the
vector $V$ of correlators.

%
%

The constraint \eqref{eq:ineq1} leads to the
following quantum-like characterization of realizable two-point correlators in any
statistical theory. 

\begin{thm}
The correlators satisfy
\begin{equation}
\label{eq:ntlm}
\begin{array}{l}
\abs{c_1 c_2 - c_3 c_4 - \mathcal{M}_{12} + \mathcal{M}_{34}} \leq \sigma_1 \sigma_2 + \sigma_3
\sigma_4 \\
\abs{c_1 c_3 - c_2 c_4 - \mathcal{M}_{13} + \mathcal{M}_{24} } \leq \sigma_1 \sigma_3 + \sigma_2
\sigma_4 \\
\abs{c_2 c_3 - c_1 c_4 - \mathcal{M}_{23} + \mathcal{M}_{14}} \leq \sigma_2 \sigma_3 + \sigma_1
\sigma_4,
\end{array}
\end{equation}
where $\sigma_i^2 = 1 - c_i^2$.
\end{thm}

\emph{Proof}.
The $4 \times 4$ matrix $\mathcal{C}$ can be partitioned into blocks as
follows

\begin{equation}
\label{eq:covmat}
\mathcal{C} =
\begin{bmatrix}
D_{12} & N \\
N^T & D_{34},
\end{bmatrix}
\end{equation}
where $D_{12}$, $N$ and $D_{34}$ are $2 \times 2$ matrices. Because
$\mathcal{C} \succeq 0$ so are $D_{12} \succeq 0$ and $D_{34} \succeq
0$. Therefore,

\begin{equation}
\det(D_{ij}) = \sigma_i^2 \sigma_j^2 - (\mathcal{M}_{ij} - c_i c_j)^2 \geq 0,
\end{equation}
namely,
\begin{equation}
\abs{\mathcal{M}_{ij} - c_i c_j} \leq \sigma_i \sigma_j,
\end{equation}
for $i,j=1,2$ and $i,j=3,4$. This together with the triangle
inequality imply

\begin{equation}
\abs{c_1 c_2 - c_3 c_4 - \mathcal{M}_{12} + \mathcal{M}_{34}} \leq
\abs{\mathcal{M}_{12} - c_1 c_2} + \abs{\mathcal{M}_{34} - c_3 c_4}
\leq \sigma_1 \sigma_2 + \sigma_3 \sigma_4.
\end{equation}
All other symmetries of this inequality in \eqref{eq:ntlm} are
obtained by swapping rows and the respective columns of
$\mathcal{C}$.\\[0.3ex]

The next corollary suggests that very little is needed to reproduce
the set of quantum mechanical two-point binary correlators.\\[0.3ex]

\begin{cor}
The correlators vector $V$ is realizable in quantum mechanics
if and only if  \eqref{eq:ineq1} holds for some positive
semi-definite matrix $\mathcal{M}$ whose diagonal entries all equal
$1$, and for which one of the terms,
$\mathcal{M}_{12} - \mathcal{M}_{34}$, \, $\mathcal{M}_{13} -
\mathcal{M}_{24}$, \, $\mathcal{M}_{23} - \mathcal{M}_{14}$, vanishes.
In such a case,

\begin{equation}
\label{eq:tlm}
\begin{array}{l}
\abs{c_1 c_2 - c_3 c_4} \leq \sigma_1 \sigma_2 + \sigma_3 \sigma_4 \\
\abs{c_1 c_3 - c_2 c_4} \leq \sigma_1 \sigma_3 + \sigma_2 \sigma_4 \\
\abs{c_2 c_3 - c_1 c_4} \leq \sigma_2 \sigma_3 + \sigma_1 \sigma_4.
\end{array}
\end{equation}
\end{cor}

The condition \eqref{eq:tlm}, which from within quantum mechanics has been shown to be necessary and
sufficient for quantum-realizable correlators independently by Tsirelson, Landau, and Masanes
\cite{L88,T87,M03}, is obtained here without assuming quantum
mechanics, but rather from a subtle restriction on the structure of $\mathcal{M}$
in any statistical theory. 

\emph{Proof}.
Suppose, for example, that $\mathcal{M}_{12} - \mathcal{M}_{34} = 0$,
in which case the first inequality in \eqref{eq:ntlm} coincides with
the first inequality in \eqref{eq:tlm}. All other symmetries of this
inequality immediately follow for they are all equivalent (upon squaring, all these inequalities become identical: $2\sigma_1^2\sigma_2^2\sigma_3^2\sigma_4^2+2c_1c_2c_3c_4+2-(c_1^2+c_2^2+c_3^2+c_4^2)\ge 0$). \\[0.3ex]

\begin{figure}[htb]
\centering

\includegraphics[width=0.5\textwidth]{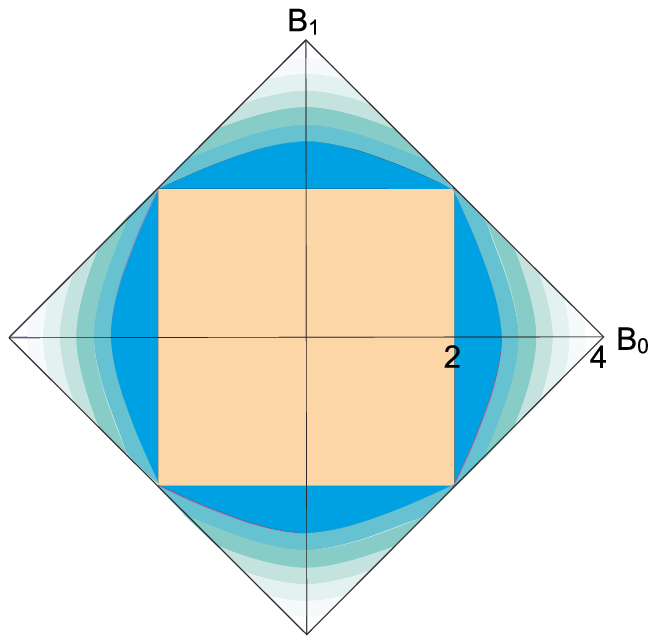}

\caption{Quantum-like bounds on any statistical theory
  \eqref{eq:ntlm}. The paler the region the larger is the difference
  $\abs{\mathcal{M}_{12} - \mathcal{M}_{34}}$. The quantum bound on
  the two-point correlators, where this difference vanishes, is shown in dark
  blue. Classical correlators make the bounded square. In this
  figure, $\mathcal{B}_x \ass c_1 + c_2 + (-1)^x (c_3 - c_4)$ is a
  symmetry of the Bell-CHSH parameter.}

\label{fig:bounds}
\end{figure}

\subsection{The covariance in quantum mechanics}

If all products can be factorized as $x_{1+i+2j} = a_i b_j$, where
$a_i$ and $b_j$ are the local outcomes of Alice and Bob per their
choices $i$ and $j$ (which actually amounts to the existence of local hidden variables), then \eqref{eq:ntlm} reduces to the set of classical correlators \cite{Fine}. The next theorem shows that when the products
are associated with operators, a similar factorization leads to the set of quantum realizable two-point binary correlators. An important difference then, between models of local hidden variables and quantum mechanics, is the non-commutativity of Alice's operators, as well as the non-commutativity of Bob's operators, which allows quantum mechanics to reach stronger correlations.

\begin{thm}
Let $X_1 \ass A_0 B_0$, $X_2 \ass A_1 B_0$, $X_3 \ass A_0
B_1$, and $X_4 \ass A_1 B_1$, where the commuting operators $A_i$ and $B_j$
are self-adjoint with $\pm 1$ eigenvalues. Then the
correlations satisfy \eqref{eq:tlm}.
\end{thm}

\emph{Proof}.
The entries,
$\mathcal{M}_{12} = \langle X_1 X_2 + X_2 X_1 \rangle / 2 = \langle\{A_0,
A_1\} \rangle / 2 = \langle X_3 X_4 + X_4 X_3 \rangle / 2  =
\mathcal{M}_{34}$, and $\mathcal{M}_{13} = \langle X_1 X_3 + X_3 X_1 \rangle / 2 = \langle\{B_0,
B_1\} \rangle / 2 = \langle X_2 X_4 + X_4 X_2 \rangle / 2 =
\mathcal{M}_{24}$. By the preceding theorem this
is all that is needed to produce the quantum set of realizable
two-point binary correlators. \\[0.3ex]

This result naturally carries over to an almost quantum theory \cite{AQ} in
which $A_i B_j |\psi\rangle = B_j A_i |\psi \rangle$ for some states, but not necessarily all of them. So in quantum theory, as well as in
an almost quantum theory the matrix $\mathcal{M}$ has both
$\mathcal{M}_{12} - \mathcal{M}_{34}$ and $\mathcal{M}_{13} -
\mathcal{M}_{24}$ vanish. Interestingly, due to the preceding theorem
there may exist theories, where only one of these terms vanishes, which
nevertheless produce the set of quantum mechanical two-point correlators.


The (almost) quantum covariance in which $\mathcal{M}_{12} -
\mathcal{M}_{34} = 0$ and $\mathcal{M}_{13} - \mathcal{M}_{24} = 0$
will henceforth be denoted as $\mathcal{C}^Q$.\\[0.3ex]

\section{Nonlocality and Tsallis entropy}

In quantum and almost quantum theories the extent of nonlocality may
be quantified by a non-additive measure of entropy.

\begin{figure}[htb]
\centering

\includegraphics[width=0.5\textwidth]{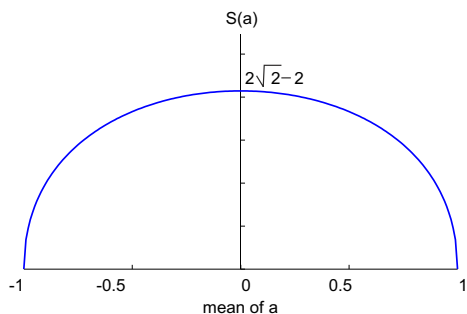}

\caption{Tsallis entropy $S(a)$ quantifies the extent of
  nonlocality in the Bell-CHSH experiment.}
\label{fig:Tsallis}
\end{figure}

\begin{thm}
In quantum and almost quantum theories
\begin{equation}
\label{eq:Tsallis}
\abs{\mathcal{B}} \leq 2 + S(a,b)
\end{equation}
where $\mathcal{B}$ is the Bell-CHSH parameter, and $S(a,b)$ is either
$S(a)$ or $S(b)$, the smallest among them, where $S(a)$ and $S(b)$ are the
Tsallis entropies \cite{Tsallis} with parameter $1/2$ of a $\pm 1$-valued random variables
$a$ and $b$ whose means are, respectively, $\langle \{A_0, A_1\} \rangle/2$ and
$\langle \{B_0, B_1\} \rangle / 2$. The right hand side in this
inequality takes values between the Bell limit,
$2$, and the Tsirelson's bound, $2 \sqrt{2}$ (see
Figure~\ref{fig:Tsallis}). The Bell bound is attained when one of the
pairs, either $A_0,A_1$ or $B_0, B_1$, commute, and the Tsirelson's
bound is attained when both anti-commute.
\end{thm}

\emph{Proof}.
The covariance matrix \eqref{eq:cv1} can be partitioned as
\begin{equation}
\mathcal{C}^Q =
\begin{bmatrix}
D_{12} & N \\
N^T & D_{34},
\end{bmatrix}
\end{equation}
where $D_{12}$, $N$ and $D_{34}$ are $2 \times 2$ matrices. Because
$\mathcal{C}^Q \succeq 0$ so are $D_{12} \succeq 0$ and $D_{34} \succeq
0$. Let $g \ass [1, \; \pm 1]$ and write

\begin{equation}
g D_{ij} g^T = 2(1 \pm \mathcal{M}_{ij}) - (\langle X_i \rangle \pm \langle X_j
\rangle)^2 \geq 0,
\end{equation}
namely,
\begin{equation}
\abs{\langle X_i \rangle \pm \langle X_j \rangle} \leq \sqrt{2(1 \pm
  \mathcal{M}_{ij})},
\end{equation}
for $i,j=1,2$ and $i,j=3.4$. This together with the triangle
inequality yield
\begin{equation}
\abs{\mathcal{B}} \leq \abs{\langle X_1 \rangle
  + \langle X_2 \rangle} + \abs{\langle X_3 \rangle - \langle X_4
  \rangle} \leq
\sqrt{2(1 + d)} + \sqrt{2(1 - d)}.
\end{equation}
where $d \ass \langle \{A_0,A_1\}\rangle / 2 = \mathcal{M}_{12} =
\mathcal{M}_{34}$. Let $y$ be a $\pm 1$-valued random variable whose
mean is $d$, i.e., $p(y = \pm 1) = (1 \pm d)/2$. The above relation
can now be written as

\begin{equation}
\abs{\mathcal{B}} \leq 2 + S(a),
\end{equation}
where the Tsallis entropy of $a$ is given by
\begin{equation}
S(a) \ass \frac{1}{q - 1} \left[ 1 - \sum_{i = \pm 1} p(a =
  i)^q. \right].
\end{equation}
with $q=1/2$.
Repeating all of the above calculations for $\tilde{\mathcal{C}}^Q$ instead of
$\mathcal{C}^Q$, where $\tilde{\mathcal{C}}^Q$ is obtained by permuting
the second and third columns of $\mathcal{C}^Q$ and then its
second and third rows, the parameter $d \ass \langle \{B_0, B_1\}
\rangle / 2 = \mathcal{M}_{13} = \mathcal{M}_{24}$.


This may strengthen different approaches, e.g. \cite{Oppenheim}, seeking for a natural relation between uncertainty and nonlocality.

Note also that quantum and almost quantum theories will generally have
different bounds in \eqref{eq:Tsallis}, depending on the pairs
$A_0,A_1$ and $B_0,B_1$.

\section{Verification using weak measurements} The above
analysis extends the ordinary Bell-CHSH experiment by introducing
$d = \langle \{A_0, A_1\} \rangle / 2$ or $d = \langle \{B_0, B_1\} \rangle / 2$, i.e. a pair of local operators at either Alice's side, Bob's side of both. In Alice's case for instance, this $d$ can be theoretically found once determining $A_0$ and $A_1$. However, one may question
the practical feasibility of inferring it with respect to the entangled state Alice and Bob share at the same run of the ordinary Bell-CHSH experiment. We propose to measure it by
employing a weak measurement \cite{AAV} of the Hermitian operator
$\{A_0, A_1\} / 2$ on Alice's side, prior to her ``strong'' projective
measurement. Weak measurement is known on theoretical \cite{Theory}
and experimental \cite{Exp} grounds to asymptotically preserve
entanglement, hence in the so called ``weak limit'' of an almost
vanishing coupling constant between Alice's qubit and the measuring
pointer, the back-action of the measurement would be negligible. When
accumulating a large enough statistics, the expectation value $d$ can
be inferred with arbitrarily high accuracy. Even though each run can be
thought of as measuring weakly a pre- and post-selected system, we can
take the weighted sum over all weak values \cite{AAV} for generating the required
expectation values. The same experimental procedure can be similarly applied to any multipartite scenario.

\section{Relation to the NPA Hierarchy}

The covariance $\mathcal{C}^Q$ is the Schur complement of
\begin{equation}
\mathcal{N} \ass
\begin{bmatrix}
1 & V^T \\
V & \mathcal{C}^Q + V V^T
\end{bmatrix}
\end{equation}
Therefore, $\mathcal{C}^Q \succeq 0$ if
and only if $\mathcal{N} \succeq 0$. This $\mathcal{N}$ may
be viewed as a symmetrization of a Hermitian matrix $\Gamma$ similar
to those used in \cite{NPA}. In particular,

\begin{equation}
\label{eq:sym}
\mathcal{N} = \frac{1}{2}\left(\Gamma + \Gamma^T\right)
\end{equation}
where $\Gamma$ is a submatrix in one of the levels of the NPA
construction.

The symmetrization \eqref{eq:sym} allows entries whose values are
otherwise inaccessible in the underlying experiment to be included in
the derived bound. In fact, terms like, e.g., $\langle
\{ A_0,A_1 \} \rangle/2$ which are missing from $\Gamma$ have been shown in
the preceding theorem to determine the extent of nonlocality. As
mentioned above, bounds involving both local and nonlocal
correlations are partly motivated by a possible application of weak
measurements.


\section{Tripartite covariance}
To examine the strength and applicability of the proposed formalism, we shall analyze in this section and in the next one two kinds of common generalizations of the Bell-CHSH setup. First, the covariance $\mathcal{C}^Q$ may be defined for any number of parties and any number
of measurement devices. In the tripartite case, for example, where Alice, Bob,
and Charlie each have a pair of measurement devices, the operators
$X_m \ass A_i B_j C_k$, \, $m=1+i+2j+4k$, where the commuting triplets
$A_i$, $B_j$, and $C_k$ are self-adjoint. Here $\mathcal{C}^Q$ is an $8 \times
8$ positive semidefinite matrix.

The tripartite covariance matrix implies bounds that
may be used to characterize the set of quantum realizable three-point
correlators, $\langle A_i B_j C_k \rangle$. In this respect the results of the
preceding theorems hold for any $4 \times 4$ submatrix of any matrix
obtained by permuting the columns and the respective rows of
$\mathcal{C}^Q$. In one case,
applying the reasoning of the last theorem
leads to a bound tighter than Mermin's inequality \cite{Mermin}
\begin{equation}
\label{eq:merm}
\abs{\langle A_0 B_0 C_0 \rangle + \langle A_1 B_1 C_0 \rangle +
  \langle A_0 B_1 C_1 \rangle - \langle A_1 B_0 C_1 \rangle} \leq
\sqrt{2(1+d)} + \sqrt{2(1-e)},
\end{equation}
where $d = \langle \{A_0 B_0, A_1 B_1\} \rangle / 2$ and $e = \langle
\{A_0 B_1, A_1 B_0\} \rangle / 2$. If both pairs,
$(A_0,A_1)$ and $(B_0,B_1)$, commute then the right hand side in
\eqref{eq:merm} equals the Bell limit, $2$. if, on the other hand,
either one of them anti-commute, in which case $d=-e$, then the right
hand side in this inequality reads $2 \sqrt{2(1+d)} \leq 4$.

The tripartite covariance may also be composed of both two- and
three-fold operators. For example, applying the first theorem to the
covariance of the operators $X_1 = A_0 B_j$, $X_2 = A_1 B_j$, $X_3 =
A_0 B_i C_k$, and $X_4 = A_1 B_i C_k$, yields

\begin{multline}
\abs{\langle A_0 B_j \rangle \langle A_1 B_j \rangle - \langle
 A_0 B_i C_k \rangle \langle A_1 B_i C_k \rangle} \leq \\
\sqrt{(1-\langle A_0 B_j \rangle^2) (1-\langle A_1 B_j \rangle^2)} +
\sqrt{(1-\langle A_0 B_i C_k \rangle^2) (1-\langle A_1 B_i C_k \rangle^2)}.
\end{multline}
which generalizes the bipartite inequality in \cite{L88,T87,M03}.
The last theorem implies in this case

\begin{equation}
\abs{\langle A_0 B_j \rangle + \langle A_1 B_j \rangle +
  \langle A_0 B_i C_k \rangle - \langle A_1 B_i C_k \rangle} \leq
2 + S(a,bc),
\end{equation}
where the means of $a$ and of $bc$ are, respectively, $\langle \{A_0,
A_1\} \rangle / 2$ and $\langle \{B_j, B_i C_k\} \rangle / 2$.

Consider now a tripartite covariance composed of only two-fold
products, e.g., $X_1 = A_0 B_j$, $X_2 = A_1 B_j$, $X_3 = A_0 C_k$, and
$X_4 = A_1 C_k$. By the last theorem

\begin{equation}
\label{eq:ineq}
\begin{array}{l}
\abs{\langle A_0 B_j \rangle + \langle A_1 B_j \rangle + \langle A_0
  C_k \rangle - \langle A_1 C_k \rangle} \leq 2 + S(a,bc) \\[0.3ex]
\abs{\langle A_i B_0 \rangle + \langle A_i B_1 \rangle + \langle B_0
  C_k \rangle - \langle B_1 C_k \rangle} \leq 2 + S(b,ac) \\[0.3ex]
\abs{\langle A_i C_0 \rangle + \langle A_i C_1 \rangle + \langle B_j
  C_0 \rangle - \langle B_j C_1 \rangle} \leq 2 + S(c,ab),
\end{array}
\end{equation}
where the means of $a$, $b$, and $c$ are, respectively, $\langle
\{A_0, A_1\} \rangle / 2$, $\langle \{B_0, B_1\} \rangle / 2$, and $\langle
\{C_0, C_1\} \rangle / 2$. Similarly, the means of $ab$, $ac$, and
$bc$ are, respectively, $\langle A_i B_j \rangle$, $\langle A_i C_k
\rangle$, and $\langle B_j C_k \rangle$. These inequalities may be
interpreted as follows. The first one, for example, suggests that the
extent of nonlocality distributed between Alice-Bob and Alice-Charlie pairs is
bounded by the local uncertainty at Alice site and also by the
uncertainty underlying the Bob-Charlie link. The greater these
uncertainties are the stronger this nonlocality may get.





\section{Further generalization of the covariance matrix}

The second kind of generalization refers to the natural case where Alice and Bob each have a two-level system, but now they can perform measurements in more than two incompatible bases (this is of course a very realistic scenario). For instance,
when Alice and Bob may each perform three different kinds of
measurements  (still having $\pm 1$ outcomes), the set of products
becomes $X_k \ass A_i B_j$, where $k=1+i+3j$, and $i,j \in \{0,1,2\}$.
Under the assumption of local realism one finds the following Bell
inequality, $\abs{\mathcal{B}'} \leq 4$, where $\mathcal{B}' = c_1 +
c_2 - c_3 + c_4 + c_5 + c_6 - c_7 + c_8$. This inequality is obtained
from the well-studied $I_{3322}$ inequality~\cite{I3322} by assuming $\pm 1$
outcomes rather than $0,1$ and by taking vanishing one-point
correlators.


Let $\mathcal{C}_{123}$ be the covariance of $X_1$, $X_2$, and
$X_3$. Similarly, let $\mathcal{C}_{456}$ and $\mathcal{C}_{78}$ be
the covariances of $X_4$, $X_5$, $X_6$, and of $X_7$, $X_8$,
respectively. Because $g^T \mathcal{C} g \geq 0$,
namely, $\abs{g^T V} \leq \sqrt{g^T \mathcal{M} g}$, for any vector
$g$, it follows that

\begin{multline}
\abs{\mathcal{B}'} \leq \abs{c_1 + c_2 - c_3} + \abs{c_4 + c_5 + c_6}
+ \abs{c_7 - c_8} \leq \\
\sqrt{g_{++-}^T \mathcal{M}_{123} g_{++-}} + \sqrt{g_{+++}^T
  \mathcal{M}_{456} g_{+++}} +
\sqrt{g_{+-}^T \mathcal{M}_{78} g_{+-}} = \\
\sqrt{3 + 2 d - 2\left(e + f\right)} + \sqrt{3 + 2 d +
2\left(e + f\right)} + \sqrt{2 - 2d},
\end{multline}
where $g_{+++}$, etc., are vectors whose entries are $1$ or $-1$,
depending on the specification. Here, $d = \langle \{ A_0, A_1 \}
\rangle / 2$, $e = \langle \{ A_0, A_2 \} \rangle / 2$, and
$f = \langle \{ A_1, A_2 \} \rangle / 2$. It is straightforward to
show that the maximum of the right hand side is
$5$, which is obtained for $e + f = 0$, and $d=1/2$. It is worthwhile
noting that this bound coincides with numerical
approximations of the bound on the original $I_{3322}$
inequality in finite-dimensional Hilbert spaces \cite{I3322num}.


\section{Conclusions}

In this work the analysis of a certain covariance matrix gave rise to a tight
characterization of binary two-point correlators in quantum mechanics and in
a general class of nonlocal theories. This formalism has further led
to a natural measure of nonlocality given by the Tsallis
entropy. Finally, we have discussed some generalizations of this
approach and derived new bounds on tripartite two- and three-point
correlators. These predictions, which often depend not only on the
correlators but also on some anti-commutators might be experimentally
tested with the aid of weak measurements \cite{AAV}, known to preserve
entanglement \cite{Theory,Exp}. That is, the nonlocal correlators can be determined as usual by performing (strong) projective measurements on the Alice and Bob sides, and at the same time weak measurements can determine the local correlators needed for the proposed bounds. As the latter involve expectation values, rather than weak values \cite{AAV}, summation over all postselections should be performed. Hence, all the above seems to be experimentally testable.

\vspace{6pt}


\acknowledgments{We thank Sandu Popescu and Paul Skrzypczyk for
  helpful comments and discussion. A.C. acknowledges support from
  Israel Science Foundation Grant No. 1723/16. E.C. was supported by the Canada
Research Chairs (CRC) Program.}

\authorcontributions{A.C. and E.C. have both written the text and
  worked out the mathemtical proofs in this paper. }

\conflictsofinterest{The authors declare no conflict of interest.}

\reftitle{References}

\end{document}